\documentstyle[12pt, amsmath,  amsfonts, epsfig, amssymb, rotating, latexsym]{article}

\begin{document}

\def\be{\begin{equation}}
\def\eps{\epsilon}
\def\cala{{\mathcal A}}
\def\calm{{\mathcal M}}
\def\calb{{\mathcal B}}
\def\calc{{\mathcal C}}
\def\calh{{\mathcal H}}
\def\cals{{\mathcal S}}
\def\caln{{\mathcal N}}
\def\calf{{\mathcal F}}
\def\calp{{\mathcal P}}

\def\rm{\mathbb{R}}
\def\qm{\mathbb{Q}}
\def\nm{\mathbb{N}}
\def\cm{\mathbb{C}}
\def\zm{\mathbb{Z}}
\def\hm{\mathbb{H}}
\def\pd{\pi/2}
\def\ua{\underline{a}}
\def\ub{\underline{b}}
\def\tr{\textrm{tr}}
\def\im{\textrm{Im}}

\def\va{{\bf a}}
\def\vb{{\bf b}}
\def\vn{{\bf n}}
\def\vv{{\bf v}}
\def\vw{{\bf w}}
\def\dpo{d_\textrm{p.o.}}
\def\ddo{d_\textrm{d.o.}}
\def\nnpq{N_q}
\def\kdz{\overline{K_2(0)}}
\def\kot{K^{\textrm{diag}}(\tau)}
%%%%%%%%%%%%%%%%%%%%%%%%%% Format de la page %%%%%%%%%%%%%%%%%%%%%%%%%%%%%%%%%%

\textwidth= 16cm
\oddsidemargin= 0.5cm
\evensidemargin=-0.5cm
\topmargin=-1cm
\textheight= 21cm

\title{\bf Periodic orbits and semiclassical form factor in barrier billiards}
\author{O. Giraud\\
Laboratoire de Physique th\'eorique, UMR 5152 du CNRS\\
Universit\'e Paul Sabatier\\
31062 TOULOUSE CEDEX 04\\
France}

\maketitle

\begin{abstract}

Using heuristic arguments based on the trace formulas, we 
analytically calculate the semiclassical two-point correlation form factor 
for a family of rectangular billiards
with a barrier of height irrational with respect to the side of the 
billiard and located at any rational position $p/q$ from the side. 
To do this, we first obtain the asymptotic density of lengths for each 
family of periodic orbits by a Siegel-Veech formula. 
The result $\kdz=1/2+1/q$ obtained for these pseudo-integrable, 
non-Veech billiards is different but not far from 
the value of $1/2$ expected for semi-Poisson statistics and 
from values of $\kdz$ obtained previously in the case of Veech billiards. 
\end{abstract}

\pagebreak

\section{Introduction}
Quantum billiards, that is, closed compact domains in the two-dimensional
Euclidean plane, are the 
simplest model of a quantum system corresponding to physical instances 
such as quantum dots
or microstructures. The statistical properties of the quantum energy 
levels of such systems have
been investigated, and it turns out that the statistical quantum 
behaviour can be related to the
classical properties of the system. It is believed that systems 
whose classical motion is chaotic
have energy levels behaving like eigenvalues of random matrix ensembles \cite{BohGiaSch84}, 
whereas the energy levels of systems whose classical motion is integrable are Poisson distributed,
i.e. they behave like independent uniformly distributed random variables \cite{BerTab77a}. Both
numerical evidence and some analytical results support these conjectures 
\cite{AndAlt95, AgaAndAlt95, Mar98}.\\
Among systems which are classically neither chaotic nor integrable, some systems have been found to 
display an eigenvalue statistics which is intermediate between the Poisson and the
Random matrix distribution. The characteristics of such intermediate statistics are
\cite{BogGerSch99} level repulsion, exponential decrease of the nearest-neighbour 
spacing distribution at infinity and linear asymptotic behaviour of the number variance
(which is related to a non-vanishing form factor at small arguments).
The form factor at the origin is equal to $1$ for classically integrable systems, to $0$
for chaotic systems, and it is found numerically to take values between 0 and 1 for intermediate
statistics, the case $\kdz=1/2$ corresponding to semi-Poisson statistics \cite{BogGerSch99}. 
Numerous quantum systems have been found to display numerically intermediate 
statistics: for example, pseudo-integrable systems such as
rational polygonal billiards (polygons in which all angles are commensurate 
with $\pi$) \cite{CasPro99}, or quantum maps \cite{GirMarOke04}.\\
An analytical approach to the study of level statistics is the 
semiclassical trace
formula, which gives an expansion of the density of
energy levels as a sum over periodic orbits \cite{BalBlo72, Gut89}, or families of 
periodic orbits in the case of integrable systems \cite{BerTab76}. For
diffractive systems, the trace formula can be modified to include diffractive orbits
contributions \cite{Sie99, BogPavSch00}. It can be argued however 
(see \cite{BogGirSch01}
for a discussion) that only the periodic orbits contribute to the semiclassical
form factor at small arguments, $\kdz$. The calculation of this quantity 
therefore only
requires to find the
periodic orbits and the areas occupied by the pencils of periodic orbits
 in a given system. Unfortunately, this is not
a simple task. For instance is is not known whether any acute triangle has a periodic 
orbit. In the case of rational polygonal billiards, it has been shown \cite{Mas90} that
the number $\caln(L)$ of periodic orbits of length less than $L$ is quadratically bounded, 
namely there exist $c_1$ and $c_2$ such that $c_1L^2\leq \caln(L)\leq c_2L^2$, but even for
general rational polygonal billiards exact asymptotics is not known.
There exist however certain specific rational polygonal billiards for which more precise
statements are known. For instance for Veech billiards \cite{Vee89, Vor96}, a special class
of rational polygonal billiards (whose stabilizer is a discrete cofinite subgroup of $SL(2, \rm)$),
precise asymptotics for $\caln(L)$ is known, and in \cite{BogGirSch01} it was possible to
calculate analytically the form factor at the origin for triangular Veech billiards.\\

This paper presents the calculation of the semiclassical form factor at the origin
for a  billiard which does not have this special Veech property, the barrier billiard. The barrier billiard
is one of the simplest pseudo-integrable billiards. It was introduced by Hannay and McCraw
\cite{HanMcc90} and consists of a rectangle $[0,a]\times[0,b]$ containing a barrier
described by the segment $\{\eps_0 a\}\times[0,\alpha b]$ with $0\leq\eps_0, \alpha<1$
(see Figure \ref{billard} left). It is a rational polygonal billiard with six angles equal to $\pi/2$
and one angle equal to $2\pi$. It is therefore a pseudo-integrable billiard 
\cite{BerRic81},
and the movement in phase space takes place on a surface of genus 2.
When the height of the barrier is such that $\alpha\in\qm$ then the barrier billiard
is a Veech billiard. But when $\alpha$ is irrational the billiard loses 
this property. 
Nevertheless, from results obtained in \cite{EskMasSch01}, 
it is still possible to work out the distribution of the 
periodic orbits in this latter case, and thus calculate analytically 
the semiclassical form 
factor at the origin,
provided the position  of the barrier is a rational number with
respect to the size of the side: $\eps_0=p/q$ with $p,q\in\nm$ coprime. 
We will first devise a method to obtain a complete 
characterization of the periodic orbit pencils in
the non-Veech barrier billiard (Section \ref{section3}). We then
rigourously derive asymptotics for each family of periodic orbit pencils
 (Section \ref{onpeutremplacer}), then use this result to calculate the 
semiclassical form factor at small arguments 
(Section \ref{calculff}). Previously obtained
analytical results show that the semi-classical form factor at the origin 
takes non-universal values between 0 and 1. For Veech triangular 
billiards with angles $(\pi/2, \pi/n, \pi/2-\pi/n)$,
the value $K_2(0)=\frac{1}{3}(n+\epsilon(n))/(n-2)$ with
$\epsilon(n)=0,2$ or $6$ was found \cite{BogGirSch01}. 
For a rectangular billiard perturbed by an
Aharonov-Bohm flux line, we obtained
$K_2(0)=1-\kappa\overline{\alpha}+4\overline{\alpha}^2$
where $\overline{\alpha}\in[0,1/2[$ is the strength of the magnetic flux 
and $\kappa$ a rational depending on the position of the flux line in the
billiard (for irrational positions, $\kappa=3$) \cite{BogGirSch01}.
For a circular billiard perturbed by an Aharonov-Bohm flux line, 
a similar result $K_2(0)=1-\kappa\overline{\alpha}(1-\overline{\alpha})$,
with $\kappa\in[0,2]$  an explicit function of the position of the flux,
was derived \cite{TheseGir02}.
In the case of the barrier billiard, we obtain $K_2(0)=1/2+1/q$.
This value depends on the position of the barrier inside the rectangle,
which reflects the fact that the structure and the properties of periodic 
orbits strongly depend on it. This analytical expression for $K_2(0)$ extends previous
results to the case of non-Veech polygonal billiards.

\section{Periodic orbits in the barrier billiard}
\label{section3}
The aim of this section is to characterize periodic orbits
in a barrier billiard. We first begin by the simple case of a
rectangular billiard.

\subsection{Periodic orbits in the rectangular billiard}
\label{casrectangle}
Let us consider a rectangle of area $\cala=a\times b$ with Dirichlet 
boundary conditions. It is easy to work out the density of the lengths 
of periodic orbits. Any orbit in the rectangle can
be unfolded into a straight line in a torus (a rectangle with periodic
 boundary conditions) of size $2a\times 2b$; a periodic orbit is 
therefore defined by two integers $M$ and $N$ and has length
\begin{equation}
\label{lprectangle}
l_p=\sqrt{(2 M a)^2+(2 N b)^2}.
\end{equation}
If we restrict ourselves to $(M,N)$ in the upper right quadrant, 
each family of periodic orbits occupies an area $4\cala$ ($2\cala$ 
for the orbit itself, $2\cala$ for its time-reverse). The number 
$\caln(l)$ of pencils of length less than $l$ is just the number of 
lattice points $(2Ma, 2Nb)$ within a (quarter of a) disk of radius $l$.
It has the asymptotic expression $\caln(l)\sim\pi l^2/16\cala$.
The corresponding density of periodic orbits is the derivative of $\caln(l)$:
\begin{equation}
\label{rhol}
\rho(l)\sim\frac{\pi l}{8\cala}.
\end{equation}
The density of primitive periodic orbits is given by (see e.g. \cite{BogGirSch01})
\begin{equation}
\label{densiterectangle}
\rho_{pp}(l)\sim\frac{3 l}{4\pi\cala}.
\end{equation}
We want to obtain a similar result for the barrier billiard. In the rest
of this section we investigate the periodic orbits of the barrier billiard,
and Section \ref{onpeutremplacer} leads to Equation \eqref{rhoppf} which
gives the density of primitive periodic orbits for the barrier billiard.

\subsection{The translation surface}
\label{transsurf}
Instead of studying directly the barrier billiard itself, we will consider the equivalent 
problem of studying the translation surface associated to this billiard \cite{GutJud00}. 
\begin{figure}[ht]
\begin{center}
\epsfig{file=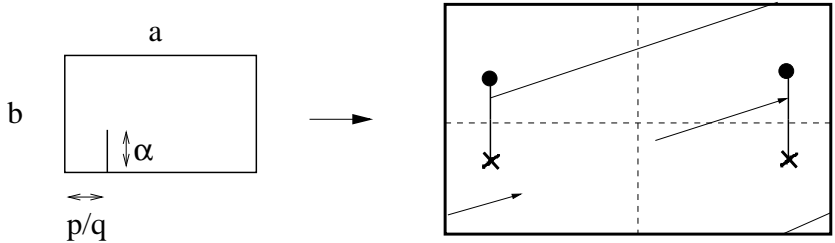,width=11cm}
\end{center}
\caption{The barrier billiard and its translation surface}
\label{billard}
\end{figure}
A construction due to Zemlyakov and Katok \cite{ZemKat76} shows that the translation surface 
associated to a generic rational polygonal billiard  is obtained by unfolding the polygon 
with respect to each of its sides, which means gluing to the initial polygon its images by reflexion 
with respect to each of its sides and repeating the operation. If the angles of the polygon are
$\alpha_i=\pi m_i/n_i$ and $N$ is the least common multiple of the $n_i$, then $2N$ copies of the initial billiard
are needed. Here all the angles are multiples of $\pi/2$,  therefore only 4 copies are needed,
and the translation surface $S$ obtained by this construction is represented 
in Figure \ref{billard} (right). In this surface, all opposite sides are identified. 
Any trajectory in the barrier billiard can be unfolded to a straight line on the translation surface.
The surface $S$ is of genus 2: there are two singular angles of measure $4\pi$ 
that we will represent respectively by $z_1$ (a dot in Figure \ref{billard}) and $z_2$ (a cross
in Figure \ref{billard}). The two singularities are traditionally
called saddles \cite{EskMasZor03} and a geodesic joining them is called a saddle-connexion.

\subsection{Periodic orbits in the barrier billiard}
\label{pobb}
In this subsection, our aim is to describe qualitatively the periodic 
orbits in the barrier billiard in a given direction.
On translation surfaces the periodic orbits occur in pencils, or 
cylinders, of periodic orbits of
same length. These cylinders are bounded by saddle-connexions and are characterized by their length 
and their height.
Let us consider a 'rational direction' on the translation surface $S$: 
\begin{equation}
\label{vecteurs}
\vv=(2 M a/q, 2 N b),
\end{equation} 
with $M$ and $N$ two coprime positive integers. 
The length of the vector $\vv$ is
\begin{equation}
\label{lgbarr}
l_p=\sqrt{(2a M/q)^2+( 2 b N)^2}.
\end{equation}
Let us label by the integers $k=0$, 1,..., $q-1$ the positions on the
translation surface such that the barrier on the ''left'' of the 
translation surface in Figure \ref{billard} be at position $p-1$
and the barrier on the ''right'' in Figure \ref{billard} be at position 
$q-p$ (see Figure \ref{unfolded}).
Since the opposite sides on the translation surface are identified, then 
when a trajectory hits the barrier at position $p-1$ it 
reappears at position $q-p$, and vice-versa. The translation by vector 
$\vv$ induces a permutation
$\sigma_{\vv}$ of the positions $\{0, 1,..., q-1\}$. 
Let us define
\begin{equation}
w_1=\min\left\{k\in{\nm}; \sigma^{k}_{\vv}(p-1)\in\{p-1, q-p\}\right\}
\end{equation}
and in the same way
\begin{equation}
w_2=\min\left\{k\in{\nm}; \sigma^{k}_{\vv}(q-p)\in\{p-1, q-p\}\right\}.
\end{equation}
A translation by the vector $w_1 \vv$ takes $z_1$ to itself and defines
a saddle-connexion of length $w_1 l_p$. The second saddle-connexion 
joining $z_1$ to itself starts at position $q-p$ and its length 
is $w_2 l_p$.\\
Figure \ref{unfolded} shows, as an example, the two saddle-connexions 
going from $z_1$ to itself in the
direction $(9,2)$ for $p/q=1/3$. The translation by the vector $\vv$ induces 
the permutation $(012)\mapsto (021)$.
One of the saddle-connexions goes from the position 0 to itself
and has a length $l_p$; the other goes from position 2 to itself 
via position 1 and has a length $2l_p$.
In any direction, there are always two saddle-connexions going from 
$z_1$ to itself, and, in the same way, two from $z_2$ to itself. 
These four saddle-connexions form the boundary of three cylinders of
periodic orbits (see Figure \ref{3cylindres}). 
\begin{figure}[ht]
\begin{center}
\epsfig{file=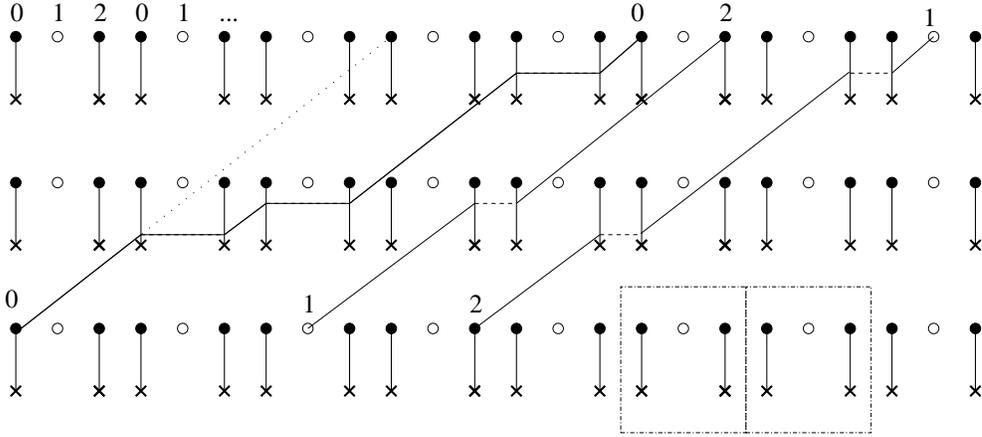,width=13cm}
\end{center}
\caption{Starting from points of abscissa 0, 1 or 2 ( for $q=3$) in the direction $(M=9,N=2)$,
one arrives at 0, 2 or 1: there are two saddle-connexions $0\to 0$ and $2\to 1\to 2$.}
\label{unfolded}
\end{figure}
The lengths of these cylinders are necessarily of the form $w_1 l_p$, $w_2 l_p$
and $(w_1+w_2)l_p$, with $w_i\in\nm$, and their heights $(2b/M)h_i$ are such that $h_1+h_3\in\zm$,
$h_2+h_3\in\zm$ and $h_3-\sigma \delta_2\in\zm$ for some $\sigma=\pm 1$ and $\delta_2=\{M\alpha\}$,
the fractional part of $M\alpha$. For instance in Figure 
\ref{3cylindres}, there is one cylinder immediately above the saddle-connexion $0\to 0$, one immediately 
above the saddle-connexion $2\to 1\to 2$, and the third cylinder is below both.
\begin{figure}[ht]
\begin{center}
\epsfig{file=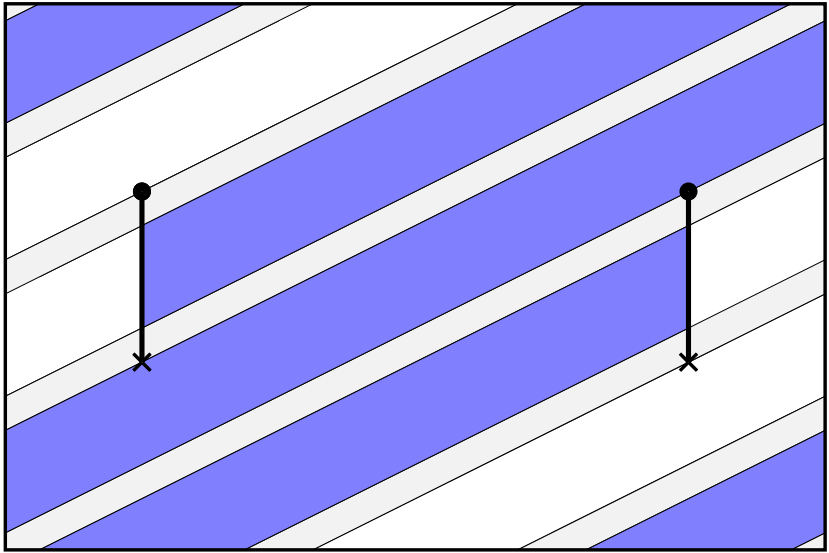,width=11cm}
\end{center}
\caption{Three cylinders of periodic orbits bounded by four saddle-connexions in the case $M=4, N=1$.}
\label{3cylindres}
\end{figure}

The results of this section can be summed up as follows. We set
\begin{eqnarray}
s_1=h_1+h_3\nonumber\\
s_2=h_2+h_3,
\end{eqnarray}
so that $s_1$ and $s_2$ are integers. Then in each direction $\vv$ defined by
$(M,N)$ with $M$ and $N$ coprime, there are three cylinders of periodic orbits
of lengths $w_i l_p$ and heights $(2b/M)h_i$ with $i=1,2,3$. 
The cylinders can be described
by the following five characteristic numbers:
\begin{itemize}
\item[-]  the integers $w_1$ and $w_2$ (giving the lengths of the two short cylinders and the 
length $(w_1+w_2)l_p$ of the long cylinder)
\item[-] the real number $h_3$ (giving the height $(2b/M)h_3$ of the long cylinder)
\item[-] the integers $s_1$ and $s_2$ (giving the heights $(2b/M)(s_1-h_3)$ 
and\\
 $(2b/M)(s_2-h_3)$ 
of the short cylinders).
\end{itemize}
Note that by definition of the $s_i$ we need to have $0<h_3<\min(s_1,s_2)$. 
Also note that the condition that the sum of the areas of the cylinders 
be $4\cala$ can be expressed as $s_1 w_1+s_2 w_2=q$.

\section{Asymptotics for the periodic orbit lengths}
\label{onpeutremplacer}
Let us define $\calf$ as the set of all 4-uples 
$(w_1,w_2,s_1,s_2)\in {(\nm^{*})}^4$
such that $(s_1, s_2)$ are coprime and $s_1 w_1+s_2 w_2=q$. We say that a 
direction $\vv$ belongs to the family $f\in\calf$ if the three 
cylinders in the direction $\vv$
have the characteristic numbers $w_1,w_2,s_1,s_2$.
 The goal of this section is to calculate, for a fixed family $f\in\calf$ and a
fixed interval $I\subset[0,\min(s_1,s_2)[$, the asymptotics 
for the number $\caln^{(q)}_{f,I}(l)$ of directions $\vv$ belonging to 
the family $f$, such that $||\vv||<l$ and such that the height $h_3$ 
of the third cylinder in the direction $\vv$ belongs to the interval $I$.

\subsection{Counting periodic orbits}
The asymptotics for the number $\caln^{(q)}(l)$ of cylinders of length less than $l$
 have been calculated in \cite{EskMasSch01}. 
These asymptotics are obtained by applying a Siegel-Veech formula
to the space $\calm_q(1,1)$ of $q$-fold coverings of the torus with two branch points and
area 1. If $V(S)$ is the set of vectors associated with cylinders of periodic orbits on a
'stable' $q$-fold torus cover $S$, then it is shown that there is a constant $\kappa(S)$
depending only on the connected component $\calm(S)$ of $\calm_q(1,1)$ containing  $S$, 
such that
\begin{equation}
\label{quadraticasymptotics1}
|V(S)\cap B(T)|\sim\pi\kappa T^2,
\end{equation}
where $|.|$ denotes the cardinal of a set and $B(T)$ is the ball of radius $T$ centered at the 
origin. The constant $\kappa$ is given by the following Siegel-Veech formula:
for any continuous compactly supported $\varphi:{\rm}^2\to\rm$,
\begin{equation}
\label{quadraticasymptotics2}
\frac{1}{\tilde{\mu}(\calm(S))}\int_{\calm(S)}\hat{\varphi}\ d\tilde{\mu}
=\kappa\int_{{\rm}^2}\varphi,
\end{equation} 
where $\hat{\varphi}$ is the Siegel-Veech transform of $\varphi$ defined by
\begin{equation}
\label{quadraticasymptotics3}
\hat{\varphi}(S)=\sum_{v\in V(S)}\varphi(v)
\end{equation}
and $\tilde{\mu}$ is the measure on $\calm_q(1,1)$ (Theorem 2.4 of \cite{EskMasSch01}).
This Theorem applies to the translation surface constructed from the barrier billiard,
provided $S$ be a stable $q$-fold torus cover, which is true only if the height $\alpha$ of the
barrier is irrational.  It is shown that in this case $\calm(S)$ is the set 
$\calp_q(1,1)\subset\calm_q(1,1)$ of primitive torus covers.
The following asymptotics are then obtained (here we have a factor 1/16 
differing from the factor in \cite{EskMasSch01} because of our conventions 
for the counting of the time-reverse partner of a periodic orbit):
\begin{equation}
\label{eskin1}
\caln^{(q)}(l)\sim c\frac{\pi l^2}{16\cala}.
\end{equation}
The constant $c$ is given by
\begin{equation}
\label{eskin2}
c=\frac{q}{\nnpq}\sum_{r|q}\mu(r)
\hspace{-.5cm}\sum_{\genfrac{}{}{0pt}{}{(s_1,s_2)=1}{s_1 u_1+s_2 u_2=q/r}}\hspace{-.5cm}
u_1 u_2 (u_1+u_2) \min(s_1,s_2)\left(\frac{1}{u_1^2}+\frac{1}{u_2^2}+\frac{1}{(u_1+u_2)^2}\right)
\end{equation}
(the gcd of $s, s'$ will be noted either $\gcd(s, s')$ or simply $(s,s')$), and
\begin{equation}
\label{nqp}
\nnpq=\sum_{r|q}\mu(r)r^2\hspace{-.5cm}\sum_{\genfrac{}{}{0pt}{}{(s_1,s_2)=1}{s_1 u_1+s_2 u_2=q/r}}
\hspace{-.5cm}u_1 u_2 (u_1+u_2) \min(s_1,s_2)
\end{equation}
(Proposition 4.14 of \cite{EskMasSch01}).
The constant $\nnpq$ is the number of primitive covers of degree $q$ of a surface of
genus 2 with 2 branch points.

\subsection{Siegel-Veech formula}
The proof leading to Equation \eqref{eskin1} can be adapted to any subset of $V(S)$
provided it varies linearly under $SL(2, \rm)$ action, i.e. provided the subset
verifies $\forall g\in SL(2, \rm)$, $V(g S)=g V(S)$ 
(see Section 2 of \cite{EskMas01} for more detail).
To obtain the asymptotics for a fixed pair $F=(f,I)$ with $f=(w_1,w_2,s_1,s_2)\in\calf$
and $I$ an interval, $I\subset[0,\min(s_1,s_2)[$, let us define 
$V_{F}(S)$ the set of vectors $\vv\in{\rm}^2$ defined by \eqref{vecteurs},
such that the triple of cylinders in the direction $\vv$ belongs to 
the family $f$, with $h_3\in I$.
Then along the same lines of the proof of Theorem 2.4 in \cite{EskMasSch01}, one
can show that when the height $\alpha$ of the barrier is irrational, the
translation surface $S$ of the barrier billiard is a stable $q$-fold torus cover and
\begin{equation}
\label{vb}
|V_F(S)\cap B(T)|\sim\pi\kappa_F T^2,
\end{equation}
where the constant $\kappa_F$ is given by the Siegel-Veech formula
\begin{equation}
\label{svf2}
\frac{1}{\tilde{\mu}(\calp_q(1,1))}\int_{\calp_q(1,1)}\hat{\varphi_F}\ d\tilde{\mu}
=\kappa_F\int_{{\rm}^2}\varphi_F,
\end{equation} 
with $\hat{\varphi_F}$ the Siegel-Veech transform 
\begin{equation}
\label{svfF}
\hat{\varphi_F}(S)=\sum_{v\in V_F(S)}\varphi(v),
\end{equation}
for some continuous compactly supported $\varphi:{\rm}^2\to\rm$.

\subsection{Asymptotics for a family of periodic orbits}
Following the steps leading from the Siegel-Veech formula 
 \eqref{quadraticasymptotics1}-\eqref{quadraticasymptotics3}
to the asymptotics \eqref{eskin1}-\eqref{eskin2} in \cite{EskMasSch01}, 
we can now derive asymptotics for
the number of cylinders in each family $(f, I)$. Recall that
$\caln^{(q)}_{f,I}(l)$ is the number of  directions $\vv$ belonging 
to a family characterized by the numbers $f=(w_1, w_2, s_1, s_2)$, 
with a height $h_3\in I$, and such that $l_p$ given by Equation 
\eqref{lgbarr} is less than $l$. (Note that $\caln^{(q)}_{f,I}(l)$ 
is a number of directions and not a number of cylinders.)
Let $\rho_{f, I}(l)$ be the corresponding density. 
According to Equation \eqref{vb}, $\caln^{(q)}_{f,I}(l)$ is proportional
to $l^2$; we define the constant $c_{f,I}$ by
\begin{equation}
\label{eskin1f}
\caln^{(q)}_{f,I}(l)\sim c_{f,I}\frac{\pi l^2}{16\cala}.
\end{equation}

The proof leading to the asymptotics for $\caln^{(q)}_{f,I}(l)$ is
essentially the same as the proof in \cite{EskMasSch01}, section 4.4,
provided we replace the counting functions of the cylinders
in \cite{EskMasSch01} by counting functions of directions in which 
the cylinders belong to the family $(f, I)$ we are interested in.
We take $\varphi$ to be the characteristic function of a disc of radius 
$\epsilon$ in ${\rm}^2$. Therefore its Siegel-Veech transform $\hat{\varphi}$,
as defined by \eqref{svfF},
counts the number of directions on $S$ in which the cylinders 
belong to family $(f,I)$ and such that $l_p<\epsilon$.
For $\epsilon$ small enough, the Siegel-Veech formula \eqref{svf2} is 
equivalent to
\begin{equation}
\label{svf3}
\pi\epsilon^2 c_{f, I}=\zeta(2)\frac{1}{\tilde{\mu}(\calp_q(1,1))}
\int_{\calp_q(1,1)}\tilde{\varphi_F}\ d\tilde{\mu},
\end{equation}
where $\zeta$ is the Riemann Zeta function and
\begin{equation}
\tilde{\varphi}_{f, I}(S)=\left\{
\begin{array}{cl}
1&\textrm{ \ \ \ if the cylinders in the horizontal direction belong}\cr
 &\textrm{to the family $f$, if $h_3\in I$ and if}\ ||\vv||<\epsilon\cr
0&\textrm{ \ \ \ otherwise.}
\end{array}
\right.
\end{equation}
We define $\chi_{f, I}:{\rm}^2\mapsto\rm$ by
\begin{equation}
\chi_{f, I}(v)=\left\{
\begin{array}{cl}
1&\textrm{ \ \ \ if the cylinders in the horizontal direction belong}\cr
 &\textrm{to the family $f$, if $h_3\in I$ and if}\ ||\vv||<\epsilon\sqrt{q}\cr
0&\textrm{ \ \ \ otherwise.}
\end{array}
\right.
\end{equation}
Following  \cite{EskMasSch01}, we parametrize $\calp_q(1,1)$ and perform 
the integration in \eqref{svf3}. Part of it can be related to the integral 
over $\chi_{f, I}$, which is $\int_{{\rm}^2}\chi_{f, I}(v)\ dv=\pi\epsilon^2 q$.
The integration yields
\begin{equation}
\int_{\calp_q(1,1)}\tilde{\varphi_F}\ d\tilde{\mu}=
\frac{\pi\epsilon^2 q}{q\zeta(2)}\sum_{r|(w_1, w_2)}
\frac{\mu(r)}{r}w_1 w_2 (w_1+w_2)|I|.
\end{equation}
From  \cite{EskMasSch01} we get $\tilde{\mu}(\calp_q(1,1))=N_q/q$, with
$\nnpq$ given by \eqref{nqp}. Equation \eqref{svf3} finally gives
\begin{equation}
\label{eskin2f}
c_{f,I}=\frac{q}{\nnpq}\sum_{r|(w_1, w_2)}\frac{\mu(r)}{r}w_1 w_2 (w_1+w_2)|I|,
\end{equation}
where $|I|$ is the length of the interval $I$.

Equation \eqref{eskin2f} shows that $c_{f,I}$ depends on $I$ only 
through its length.
Is is therefore convenient to introduce the density of 
directions $p=(M,N)$ corresponding to a 
family $f$ and such that $l_p<l$ and $h_3=h$:
\begin{equation}
\label{eskin1fh}
\caln^{(q)}_{f,h}(l)\sim c_{f,h}\frac{\pi l^2}{16\cala},
\end{equation}
with $c_{f,h}$ given by
\begin{equation}
\label{eskin2fh}
c_{f,h}=\frac{q}{\nnpq}\sum_{r|q}\frac{\mu(r)}{r}w_1 w_2 (w_1+w_2)\theta_f(h)
\end{equation}
for any family $f=(w_1, w_2, s_1, s_2)$ of $\calf$ and $h\in \rm$.
The function $\theta_f$ is the characteristic function of the interval 
$[0,\min(s_1,s_2)[$.
The density of primitive periodic orbit lengths for the family 
$f\in\calf$ and $h\in\rm$ 
is
\begin{equation}
\label{rhoppf}
\rho_{pp,f, h}(l)\sim c_{f,h} \frac{3 l}{4\pi\cala}.
\end{equation}
It is easy to verify that the expression \eqref{eskin1fh} of 
$\caln^{(q)}_{f,h}(l)$ 
is consistent with the total number $\caln^{(q)}(l)$ of pencils of 
periodic orbits with length less than  $l$. This comes from the fact 
that any  pencil of periodic orbits contributing to 
$\caln^{(q)}(l)$ belongs to a certain family $f$ and has a length 
$w_i l_{p}\leq l$, which implies that $l_{p}\leq l/w_i$. Therefore 
\begin{eqnarray}
\caln^{(q)}(l)&=&\sum_{f\in\calf}\int dh\sum_{i=1}^{3}
\caln^{(q)}_{f,h}(l/w_i)\nonumber\\
&\sim&\sum_{f\in\calf}\int dh\ c_{f,h}\frac{\pi l^2}{16\cala}
\left(\frac{1}{w_1^2}+\frac{1}{w_2^2}+\frac{1}{(w_1+w_2)^2}\right).
\end{eqnarray}
Using Equation \eqref{eskin2fh}, we obtain, after integration over $h$,
\begin{eqnarray}
\caln^{(q)}(l)&\sim&\frac{\pi l^2}{16\cala}
\frac{q}{\nnpq}\sum_{\genfrac{}{}{0pt}{}{(s_1,s_2)=1}{s_1 w_1+s_2 w_2=q}}
\sum_{r|(w_1,w_2)}\frac{\mu(r)}{r} w_1 w_2 (w_1+w_2)\nonumber\\
&&\times\min(s_1,s_2)
\left(\frac{1}{w_1^2}+\frac{1}{w_2^2}+\frac{1}{(w_1+w_2)^2}\right).
\end{eqnarray}
Making the substitution  $w_i=r u_i$ and inverting the two sums, we get 
exactly the expression given by Equations (\ref{eskin1}) and (\ref{eskin2}).

\section{Calculation of the form factor at $\tau=0$}
\label{calculff}

\subsection{Definitions}
\label{section2}
The spectrum $\{E_n, n\in\nm\}$ of a quantum billiard can be described 
by the density
\begin{equation}
d(E)\equiv\sum_n \delta(E-E_n).
\end{equation}
The two-point correlation form factor is defined as the Fourier 
transform of the two-point correlation function of the density of states:
\begin{equation}
\label{formfactor}
K_2(\tau)=\int_{-\infty}^{\infty}\frac{d\eps}{\bar{d}}\langle
d(E+\eps/2)d(E-\eps/2)\rangle_{\textrm{c}} 
e^{2 i \pi \bar{d} \tau\eps}.
\end{equation}
Here the product of the densities is averaged over an energy window of width 
$\Delta E\gg 1/\bar{d}$ centered around $E=k^2$ and such that 
$\Delta E \ll E$. If $\cala$ is the area of the billiard, 
$\bar{d}=\cala/4\pi$ is the non-oscillating part of the density of states.
The subscript c means that one only considers the connected part of the
correlation function. It can be argued that in the case of 
pseudo-integrable systems, the leading term of the 
semiclassical expansion of $K_2(\tau)$ at small argument ($\tau\to 0$) 
is given in the diagonal approximation by the contribution of periodic orbits
only: $K_2(\tau)=\kot+O(\tau)$, with
\begin{equation}
\kot=\frac{1}{8\pi^2\bar{d}}
\sum_{p} \frac{|S_{p}|^2}{l_p}
\delta (l_{p}-4\pi k \bar{d} \tau)
\label{k0}
\end{equation}
(see \cite{BogGirSch01} for the derivation of this expression, based
on heuristic arguments). The sum is 
performed over all pencils of periodic orbits $p$ of length $l_p$. 
In general, there can be several pencils having exactly the same 
length: in Equation \eqref{k0}, $S_p$ is the sum of the areas 
occupied by all pencils having, when (possibly) multiply repeated, 
a length $l_p$. 
The aim of the present section is to calculate the 
semiclassical form factor at small arguments \eqref{k0},
using the result (\ref{rhoppf}) for the distribution of pencils of 
periodic orbits in the barrier billiard. Let us take a $C^{\infty}$,
compactly supported test function and integrate the distribution $\kot$
over $\tau$. If the density of periodic pencils depends linearly on $l$
(as is the case for the barrier billiard or the rectangular billiard), the
integration over families of periodic orbits yields 
$\kot=\lambda\Theta(\tau)$, where $\lambda$ is a constant
and $\Theta$ the Heaviside step function. In such a case, we define 
$\kdz=\lambda$. 

As an introduction, we first
deal with the simpler case of a rectangular billiard.

\subsection{Rectangular billiard}
In the case of the rectangular billiard, discussed in section 
\ref{casrectangle}, the periodic orbits have lengths $n l_{pp}$,
where $l_{pp}$ is given by \eqref{lprectangle} with $(M,N)$ coprime,
and $n\in\nm$ is the repetition number. The area of each pencil 
of primitive periodic orbits $pp$ is $A_{pp}=4\cala$. 
When the sides  $a$ and $b$ of the rectangle are incommensurable, there
is only one pencil of length $l_{pp}$ and therefore in Equation (\ref{k0})
$S_p=4\cala$. Equation (\ref{k0}) becomes
\begin{equation}
\kot=\frac{1}{8\pi^2\bar{d}}
\sum_{pp}\sum_{n}\frac{|4\cala|^2}{n^2 l_{pp}}
\delta (l_{pp}-4\pi k \bar{d} \tau/n)
\end{equation}
hence (using the fact that $\bar{d}=\cala/4\pi$ and turning the sum over
$pp=(M,N)$ with $M$ and $N$ coprime into an integral over $l$ with density 
$\rho_{pp}(l)$)
\begin{equation}
\label{k2rec}
\kot=\frac{8\cala}{\pi}\sum_n
\int_{0}^{\infty}dl\ \frac{1}{n^2 l}\rho_{pp}(l)
\delta (l-4\pi k \bar{d} \tau/n)
\end{equation}
The density $\rho_{pp}(l)$ of periodic orbits is given by Equation 
(\ref{densiterectangle}) and yields $\kot=1$, as expected for 
integrable systems.

\subsection{Barrier billiard}
In the case of the barrier billiard, the periodic orbits 
have a length of the form $n w l_p$ with $l_p$ given by \eqref{lgbarr}: here
the primitive length is $w l_p$ and $n$ is the repetition number. 
Two pencils of periodic orbits $p$ and $p'$ have the same length provided 
there exist repetition numbers $n$ and $n'$ such that
 $n w l_{p}=n' w' l_{p'}$. When $a$ and $b$ are incommensurable, 
this implies $p=p'$, i.e. two pencils can 
have same length only if they are in the
same direction. For a given direction $(M,N)$ with $M$ and $N$ 
coprime (which will now be labeled by $p$), 
there are three cylinders of area $\cala_i$ and 
length $w_i l_{p}$, $1\leq i \leq 3$, and therefore 
$w,w'$ belong to the set $\{w_1, w_2, w_1+w_2\}$. Equation (\ref{k0}) becomes
\begin{equation}
\label{dpo2barr}
\kot=\frac{1}{8\pi^2\bar{d}}
\sum_{p}\sum_{n}\frac{|S_{p,n}|^2}{n l_p}
\delta (n l_{p}-4\pi k \bar{d} \tau)
\end{equation}
where $l_p$ is given by \eqref{lgbarr} and $S_{p,n}$ is the sum over 
the $\cala_i$ corresponding to a $w_i$ which divides $n$:
\begin{equation}
\label{sppn}
S_{p,n}=\sum_{i=1}^{3}\cala_i \delta_{w_i|n},
\end{equation}
with $\delta_{r|t}=1$ if $r$ divides $t$, 0 otherwise.
Each area $\cala_i$ is equal to $(2b/M)h_i\times (w_i l_{p})\cos\varphi_{p}$ ($\varphi_{p}$
is the angle between the orbit and the horizontal). This can be rewritten as
\begin{equation}
\label{area}
\cala_i=\frac{4\cala}{q}h_i w_i
\end{equation}
(note that since $\sum_ih_iw_i=s_1w_1+s_2 w_2=q$, one has $\sum_i\cala_i=4\cala$, i.e. the total area
of the translation surface, as expected). Therefore $S_{p,n}$ only depends of the five numbers
$f=(w_1, w_2,s_1,s_2)$ and $h_3$, and can be rewritten:
\begin{equation}
S_{f,h_3,n}=\frac{4\cala}{q}\left[(s_1-h_3)w_1\delta_{w_1|n}+(s_2-h_3)w_2\delta_{w_2|n}
+h_3(w_1+w_2)\delta_{(w_1+w_2)|n}\right].
\end{equation}
The sum (\ref{dpo2barr}) over all periodic orbits can  be 
partitioned into sums running over primitive pencils of periodic orbits $p(f,h)$ belonging to
a family $f$ with a height of the long cylinder in $[h, h+dh[$; \eqref{dpo2barr}
becomes
\begin{equation}
\kot=\sum_{f\in\calf}\int dh\sum_{p(f,h)}\sum_{n}
\frac{|S_{p,n}|^2}{8\pi^2 n^2 l_{p} \bar{d}}
\delta(l_{p}-\frac{4\pi k \bar{d} \tau}{n}).
\end{equation}
Each of the sums corresponding to a family $f$
can be replaced, as in (\ref{k2rec}), by an integral with density 
$\rho_{pp,f,h}(l)$, and
$S_{p,n}$ by $S_{f,h,n}$:
\begin{equation}
\kot=\sum_{f\in\calf}\int dh\sum_{n}
\frac{|S_{f,h,n}|^2}{8\pi^2 n^2\bar{d}}
\int_{0}^{\infty}dl\ \frac{\rho_{pp,f,h}(l)}{l}
\delta(l-\frac{4\pi k \bar{d} \tau}{n}).
\end{equation} 
Replacing the density $\rho_{pp,f,h}$ by its expression (\ref{rhoppf}), the 
integration over $l$ becomes straightforward and yields $\kot=\kdz\Theta(\tau)$, 
where
\begin{eqnarray}
\kdz=\frac{1}{q^2}\frac{6}{\pi^2}\sum_{n}\frac{1}{n^2}\sum_{f\in\calf}\int dh
\left[(s_1-h)w_1\delta_{w_1|n}\right.\\
\nonumber 
+\left.(s_2-h)w_2\delta_{w_2|n}
+h(w_1+w_2)\delta_{(w_1+w_2)|n}\right]^2 c_{f,h}
\end{eqnarray} 
(we have used the fact that $\bar{d}=\cala/4\pi$). Expanding the square, we can perform the
summation over $n$, using the identity
\begin{equation}
\sum_{n=1}^{\infty}\frac{\delta_{w_1|n}\delta_{w_2|n}}{n^2}=\frac{\pi^2}{6}
\frac{\gcd(w_1,w_2)^2}{w_1^2 w_2^2}.
\end{equation}
The form factor can therefore be written, after simplifications using the fact that 
$\gcd(w_1,w_1+w_2)=\gcd(w_2,w_1+w_2)=\gcd(w_1,w_2)$, as
\begin{eqnarray}
\kdz&=&\frac{1}{q^2}\sum_{f\in\calf}\int dh\ c_{f,h}
\left[3h^2-2\left(s_1+s_2+q\frac{\gcd(w_1,w_2)^2}{w_1 w_2 (w_1+w_2)}\right)h\right.\nonumber\\
&+&\left.s_1^2+s_2^2+2 s_1 s_2\frac{\gcd(w_1,w_2)^2}{w_1 w_2}\right].
\end{eqnarray} 
Replacing the weight $c_{f,h}$ by its expression (\ref{eskin2fh}), we can easily perform the 
integration over $h$, which consists of terms of the form
\begin{equation}
\int_{0}^{\min(s_1, s_2)}dh\ h^{\nu}=\frac{\min(s_1,s_2)^{\nu+1}}{\nu+1}
\end{equation}
for $\nu=0,1,2$. The form factor becomes
\begin{eqnarray}
\kdz&=&\frac{1}{q \nnpq}\sum_{\genfrac{}{}{0pt}{}{(s_1,s_2)=1}{s_1 w_1+s_2 w_2=q}}
\sum_{r|(w_1,w_2)}\frac{\mu(r)}{r} w_1 w_2 (w_1+w_2)
\left[\min(s_1,s_2)^3\right.\nonumber\\
&-&\left(s_1+s_2+q\frac{\gcd(w_1,w_2)^2}{w_1 w_2 (w_1+w_2)}\right)\min(s_1,s_2)^2\nonumber\\
&+&\left.\left(s_1^2+s_2^2+2 s_1 s_2\frac{\gcd(w_1,w_2)^2}{w_1 w_2}\right)\min(s_1,s_2)\right].
\end{eqnarray} 
This sum can be evaluated with some cumbersome arithmetic manipulations; the calculation is given in 
the Appendix, and the final result is unexpectedly simple:
\begin{equation}
\label{resultatfinal}
\kdz=\frac{1}{2}+\frac{1}{q}.
\end{equation}
There are several comments to make concerning this value. First, it is close 
to the result corresponding to semi-Poisson statistics $\kdz=1/2$ 
\cite{BogGirSch01}. This result is not 
valid for $q=2$, since in that case there is an additional symmetry in the billiard,
with respect to the barrier, and the spectrum has to be desymmetrized. The calculation in
this case has been done in \cite{Wie02} for  a height of the barrier equal to $b/2$ (half the height of
the rectangle), and yields $\kdz=1/2$.
The calculation for $q=2$ and a barrier with any height has been done in \cite{TheseGir02} using
a different method, and also yields $\kdz=1/2$.

The result (\ref{resultatfinal}) is similar to previously obtained results
\cite{BogGirSch01} for rational polygonal billiards having the Veech property.
For instance for triangular billiards with angles $(\pi/2, \pi/n, \pi/2-\pi/n)$
the form factor at the origin was found to be between $1/3$ and $3/5$ \cite{BogGirSch01}.
Here the form factor lies between $1/2$ and $5/6$, which again is close to the
semi-Poisson result.

\section*{Acknowledgments}
The author thanks Professor Alex Eskin for helpful discussions.
The funding of the Leverhulme trust and the Department of Physics of
the University of Bristol, where most of this work has been done,
 are gratefully acknowledged for their support. The funding of
post-doctoral CNRS fellowship and the theoretical physics laboratory
of the University of Toulouse have made the completion of this work 
possible.

%%%%%%%%%%%%%%%%%%%%%%%%%%%%%%%%%%%%%%%%%%%%%%%%%%%%%%%%%%%%%%%%%%%%%%%%%%%%%
%                                                                           %
%                                   APPENDIX                                %     
%                                                                           %
%%%%%%%%%%%%%%%%%%%%%%%%%%%%%%%%%%%%%%%%%%%%%%%%%%%%%%%%%%%%%%%%%%%%%%%%%%%%%

\section*{Appendix}
In this appendix, we want to evalute the quantity
\begin{equation}
\label{depart}
\kdz=\frac{1}{q \nnpq}\sum_{\genfrac{}{}{0pt}{}{(s_1,s_2)=1}{s_1 w_1+s_2 w_2=q}}
\sum_{r|(w_1,w_2)}\frac{\mu(r)}{r}f(s_1, s_2, w_1, w_2,q),
\end{equation}
where
\begin{eqnarray}
\label{homogeneite}
f(s_1, s_2, w_1, w_2,q)&=& w_1 w_2 (w_1+w_2)\left[\min(s_1,s_2)^3\right.\nonumber\\
&-&\left.\left(s_1+s_2+q\frac{\gcd(w_1,w_2)^2}{w_1 w_2 (w_1+w_2)}\right)\min(s_1,s_2)^2\right.\nonumber\\
&+&\left.\left(s_1^2+s_2^2+2 s_1 s_2\frac{\gcd(w_1,w_2)^2}{w_1 w_2}\right)\min(s_1,s_2)\right].
\end{eqnarray} 
The function $f$ is homogeneous, in the sense that it verifies
\begin{equation}
f(s_1, s_2,\lambda w_1,\lambda w_2, \lambda q)=f(\lambda s_1,\lambda s_2,w_1, w_2,\lambda q)
=\lambda^3 f( s_1, s_2,w_1,w_2,q).
\end{equation}
In (\ref{depart}), the first sum goes over all integers $w_i\geq 1$ and $s_i\geq 1$, $i=1,2$, 
verifying $s_1 w_1+s_2 w_2=q$ and $\gcd(s_1,s_2)=1$. The number $\nnpq$ is given by (\ref{nqp}).
The first step is to exchange the sum over $(s_i, w_i)$ and the sum over $r$ in (\ref{depart}), and 
substitute $w_i=r u_i$: using the homogeneity of $f$, we get
\begin{equation}
\label{stari}
\kdz=\frac{1}{q \nnpq}\sum_{r|q}\mu(r)r^2
\sum_{\genfrac{}{}{0pt}{}{(s_1,s_2)=1}{s_1 u_1+s_2 u_2=q/r}}f(s_1, s_2, u_1, u_2, \frac{q}{r}),
\end{equation}
To get rid of the co-primality condition on $(s_1, s_2)$
we use the exclusion-inclusion principle, which for any function $\varphi$ gives
\begin{equation}
\label{excluinclu}
\sum_{(s,s')=1}\varphi(s,s')=\sum_{s, s'=1}^{\infty}\sum_{t=1}^{\infty}\mu(t)\varphi(t s, t s').
\end{equation}
This allows to rewrite the form factor as
\begin{eqnarray}
\kdz&=&\frac{1}{q \nnpq}\sum_{r|q}\mu(r)r^2\sum_{t=1}^{\infty}\mu(t)
\hspace{-.5cm}\sum_{t s_1 u_1+t s_2 u_2=q/r}f(t s_1, t s_2,u_1, u_2,  \frac{q}{r})\hspace{-.5cm}\\
&=&\frac{1}{q \nnpq}\sum_{r|q}\mu(r)r^2\sum_{t|q}\mu(t)t^3
\hspace{-.5cm}\sum_{s_1 u_1+s_2 u_2=q/(rt)}f(s_1, s_2, u_1, u_2, \frac{q}{r t})\hspace{-.5cm}\nonumber
\end{eqnarray}
Here the sum over $t$ from 1 to $\infty$ has been replaced by a sum over $t|q$ since for all the other
values of $t$ there is no value of $(s_1, s_2,u_1, u_2)$ fulfilling the condition $t s_1 u_1+t s_2 u_2=q/r$.
Again, the homogeneity of $f$ (Equation (\ref{homogeneite})) has been used. Setting $d=r t$
we get
\begin{eqnarray}
\label{starf}
\kdz=\frac{1}{q \nnpq}\sum_{d|q}\left(\sum_{t|d}\mu(t)\mu(\frac{d}{t}) d^2 t\right)
\sum_{s_1 u_1+s_2 u_2=q/d}f(s_1, s_2, u_1, u_2, \frac{q}{d}).\nonumber
\end{eqnarray}
We need to evaluate
\begin{equation}
\label{k0total}
\kdz=\frac{1}{q \nnpq}\sum_{d|q}\left(\sum_{t|d}\mu(t)\mu(\frac{d}{t}) d^2 t\right)(G_{q/d}+H_{q/d}),
\end{equation}
where
\begin{eqnarray}
G_n&\equiv&\hspace{-.5cm} \sum_{s_1 u_1+s_2 u_2=n}\hspace{-.5cm}u_1 u_2 (u_1+u_2)\left[\min(s_1,s_2)^3\right.\\
&-&\left.\left(s_1+s_2\right)\min(s_1,s_2)^2+\left(s_1^2+s_2^2+2 s_1 s_2\right)\min(s_1,s_2)
\right]\nonumber
\end{eqnarray}
and 
\begin{eqnarray}
H_n&\equiv&\hspace{-.5cm} \sum_{s_1 u_1+s_2 u_2=n}\hspace{-.5cm}u_1 u_2 (u_1+u_2)
\left[-q\frac{\gcd(u_1,u_2)^2}{u_1 u_2 (u_1+u_2)}\min(s_1,s_2)^2\right.\nonumber\\
&&\hspace{3cm}+\left.2 s_1 s_2\frac{\gcd(u_1,u_2)^2}{u_1 u_2}\min(s_1,s_2)\right].
\end{eqnarray} 
The quantities $G_n$ and $H_n$ will be evaluated separately.
This evaluation will require the use of a theorem proved in 
\cite{HuaOuSpeWil02}:\\
{\bf Theorem.} Let $f:\zm$$^{4}\rightarrow \cm$ such that
\begin{equation}
f(a,b,x,y)-f(x,y,a,b)=f(-a, -b, x,y)-f(x,y,-a,-b)
\end{equation}
for all integers $a,b,x$ and $y$. Then for $n\in\nm$, $n\geq 1$,
\begin{eqnarray}
\label{theoreme}
\sum_{\genfrac{}{}{0pt}{}{a,b,x,y\geq 1}{a x+b y=n}}
\left[f(a,b,x,-y)-f(a,-b,x,y)+f(a,a-b,x+y,y)\right.\nonumber\\
-\left. f(a,a+b,y-x,y)+f(b-a, b, x, x+y)-f(a+b, b, x, x-y)\right]\nonumber\\
=\sum_{d|n}\sum_{x=1}^{d-1}\left[
f(0, \frac{n}{d}, x, d)+f(\frac{n}{d},0,d, x)+f(\frac{n}{d},\frac{n}{d},d-x,-x)\right.\nonumber\\
-\left. f(x, x-d, \frac{n}{d},\frac{n}{d})-f(x,d,0,\frac{n}{d})-f(d,x,\frac{n}{d},0)\right].
\end{eqnarray}

\subsection*{a. Evaluation of $G_n$}

We can immediately point out that the identity
\begin{equation}
\min(s_1,s_2)^2-(s_1+s_2)\min(s_1,s_2)=-s_1 s_2,
\end{equation}
valid for any integers $s_1$ and $s_2$, allows to simplify $G_n$. We now need to evaluate
the sum
\begin{equation}
G_n=\hspace{-.5cm}\sum_{s_1 u_1+s_2 u_2=n}\hspace{-.5cm}
 u_1 u_2 (u_1+u_2)\min(s_1,s_2)\left(s_1^2+s_2^2-s_1 s_2\right).
\end{equation}
for any integer $n$. Writing $\min(a,b)=\frac{1}{2}(a+b-|a-b|)$, we have
\begin{eqnarray}
\label{gn2sommes}
G_n&=&\frac{1}{2}\sum_{s_1 u_1+s_2 u_2=n}\hspace{-.5cm}u_1 u_2\left[
s_1^3 u_2+s_2^3 u_1-\frac{1}{3}(s_1^3 u_1+s_2^3 u_2)\right.\\
&-&\left.\vphantom{\frac{1}{2}}(u_1+u_2)(s_1^2+s_2^2-s_1 s_2)|u_1-u_2|\right]
+\frac{2}{3}\hspace{-.5cm}\sum_{s_1 u_1+s_2 u_2=n}\hspace{-.5cm}u_1 u_2(s_1^3 u_1+s_2^3 u_2).\nonumber
\end{eqnarray}
The first sum in (\ref{gn2sommes}) can be evaluated by applying Theorem (\ref{theoreme}) to the function
\begin{equation}
f(a,b,x,y)=\frac{1}{3}\left(x y-\frac{|x y|}{2}\right)\left|(a-b)(x-y)\right|(a^2+b^2-a b)
\end{equation}
and is equal to
\begin{equation}
\frac{n^2 (n-1)}{18}\sum_{d|n}d.
\end{equation}
The second sum in (\ref{gn2sommes}) can be evaluated by applying Theorem (\ref{theoreme}) to the function
$f(a,b,x,y)=b^2y^4-b^2 x y^3$ (see \cite{HuaOuSpeWil02}). It gives
\begin{equation}
\frac{4}{3}\sum_{a x+b y=n}\hspace{-.3cm}a^3 x^2 y
=\frac{n^2}{18}\sum_{d|n}\left(3d^3+(1-4n)d\right).
\end{equation}
Finally we get
\begin{equation}
\label{gn}
G_n=\frac{n^2}{6}\sum_{d|n}\left(d^3-n d\right).
\end{equation}
If we now evaluate the quantity
\begin{eqnarray}
\sum_{s_1 u_1+s_2 u_2=n}\hspace{-.5cm}
 u_1 u_2 (u_1+u_2)\min(s_1,s_2)=\frac{1}{2}\sum_{s_1 u_1+s_2 u_2=n}\hspace{-.5cm}u_1 u_2
\left[s_1 u_2+s_2 u_1\vphantom{\frac{1}{2}}\right.\nonumber\\
\left.-\frac{1}{3}(s_1 u_1+s_2 u_2)
-(u_1+u_2)|u_1-u_2|\right]
+\frac{2n}{3}\hspace{-.5cm}\sum_{s_1 u_1+s_2 u_2=n}\hspace{-.5cm}u_1 u_2,
\end{eqnarray}
the first sum is given by Theorem (\ref{theoreme}) applied to the function
\begin{equation}
f(a,b,x,y)=\frac{1}{3}\left(x y-\frac{|x y|}{2}\right)\left|(a-b)(x-y)\right|
\end{equation}
and the second one is given by Theorem (\ref{theoreme}) applied to the 
function\\
$f(a,b,x,y)=n x y/3$; altogether, this gives
\begin{equation}
\sum_{s_1 u_1+s_2 u_2=n}\hspace{-.5cm}
 u_1 u_2 (u_1+u_2)\min(s_1,s_2)=\frac{n}{3}\sum_{d|n}\left(d^3-n d\right).
\end{equation}
Together with Equation (\ref{gn}) we get
\begin{equation}
G_{n}=\frac{n}{2}\sum_{s_1 u_1+s_2 u_2=n}\hspace{-.5cm}u_1 u_2 (u_1+u_2)\min(s_1,s_2).
\end{equation}

\subsection*{b. Evaluation of $H_n$}
We want to evaluate
\begin{eqnarray}
\label{departH}
H_n&=&\hspace{-.5cm}\sum_{s_1 u_1+s_2 u_2=n}\hspace{-.5cm} u_1 u_2 (u_1+u_2)\min(s_1,s_2)\hspace{4cm}\nonumber\\
&&\hspace{2cm}\left(-n\frac{\min(s_1,s_2)}{ u_1 u_2 (u_1+u_2)}+\frac{2 s_1 s_2}{u_1 u_2}\right)
\gcd(u_1, u_2)^2
\end{eqnarray}
for any integer $n$. Summing over all the possible values $r$ of the $\gcd$ of $u_1$ and $u_2$, and
substituting $u_i=r v_i$, we have
\begin{equation}
H_n=\sum_{r|n}\hspace{-.2cm}
\sum_{\genfrac{}{}{0pt}{}{(v_1,v_2)=1}{s_1 v_1+s_2 v_2=n/r}}\hspace{-.2cm}
r^3 v_1 v_2 (v_1+v_2)\min(s_1,s_2)
\left(-\frac{n}{r}\frac{\min(s_1,s_2)}{v_1 v_2 (v_1+v_2)}+\frac{2 s_1 s_2}{v_1 v_2}\right).
\end{equation}
Then, as before, the co-primality condition can be expressed by a sum over $t$ (see
Equation (\ref{excluinclu})). Restricting the sum over $t$ as before, 
we get 
\begin{eqnarray}
H_n&=&\sum_{r|n}\sum_{t|n}\mu(t)r^3 t  \hspace{-.5cm}\sum_{s_1 v_1+s_2 v_2=n/(r t)}\hspace{-.5cm}
 v_1 v_2 (v_1+v_2)\min(s_1,s_2)\hspace{2cm}\nonumber\\
&&\hspace{5cm}\left(-\frac{n}{r t}\frac{\min(s_1,s_2)}{v_1 v_2 (v_1+v_2)}+\frac{2 s_1 s_2}{v_1 v_2}\right).
\end{eqnarray}
Setting $d=rt$ we get
\begin{eqnarray}
\label{arriveeH}
H_n&=&\sum_{d|n}\left(\sum_{t|d}\mu(t)\frac{d^3}{t^2}\right)\sum_{s_1 v_1+s_2 v_2=n/d}\hspace{-.5cm}
 v_1 v_2 (v_1+v_2)\min(s_1,s_2)\hspace{2cm}\nonumber\\
&&\hspace{5cm}\left(-\frac{n}{d}\frac{\min(s_1,s_2)}{v_1 v_2 (v_1+v_2)}+\frac{2 s_1 s_2}{v_1 v_2}\right).
\end{eqnarray}
Let us now evaluate, for any integer $m$, the quantity
\begin{eqnarray}
K_m&=&\sum_{s_1 v_1+s_2 v_2=m}v_1 v_2 (v_1+v_2)\min(s_1,s_2)
\left(-m\frac{\min(s_1,s_2)}{v_1 v_2 (v_1+v_2)}+\frac{2 s_1 s_2}{v_1 v_2}\right)\nonumber\\
&=&-m\hspace{-.5cm}\sum_{s_1 v_1+s_2 v_2=m}\hspace{-.5cm}\min(s_1,s_2)^2
+2\hspace{-.5cm}\sum_{s_1 v_1+s_2 v_2=m}\hspace{-.5cm}s_1 s_2 (v_1+v_2)\min(s_1,s_2).
\end{eqnarray}
Let
\begin{eqnarray}
L_m&=&\hspace{-.5cm}\sum_{s_1 v_1+s_2 v_2=m}\hspace{-.5cm}(2 s_1 s_2-v_1 v_2)(v_1+v_2)\min(s_1,s_2)\\
&=&\hspace{-.5cm}\sum_{s_1 v_1+s_2 v_2=m}\hspace{-.5cm}
(2 v_1 v_2(s_1+s_2)\min(v_1,v_2)-v_1 v_2(v_1+v_2)\min(s_1,s_2))\nonumber
\end{eqnarray}
after exchanging $(s_1, s_2)$ and $(v_1, v_2)$ in the first half of the right member.
Writing $\min(a,b)=\frac{1}{2}(a+b-|a-b|)$ and applying Theorem (\ref{theoreme}) to
the function
\begin{equation}
f(a,b,x,y)=-\frac{1}{2}\left|a b (a-b)(x-y)\right|
\end{equation}
one gets
\begin{equation}
L_m=m\sum_{d|m}\sum_{x=1}^{d-1}x(d-x).
\end{equation}
Applying Theorem (\ref{theoreme}) to
the function
\begin{equation}
f(a,b,x,y)=\frac{a b-\left|a b\right|}{2}
\end{equation}
one gets
\begin{equation}
\sum_{s_1 v_1+s_2 v_2=m}\hspace{-.5cm}\min(s_1,s_2)^2=\sum_{d|m}\sum_{x=1}^{d-1}x(d-x).
\end{equation}
This proves that
\begin{equation}
K_m=\hspace{-.5cm}\sum_{s_1 v_1+s_2 v_2=m}\hspace{-.5cm}v_1 v_2(v_1+v_2)\min(s_1,s_2)
\end{equation}
and therefore
\begin{equation}
H_n=\sum_{d|n}\left(\sum_{t|d}\mu(t)\frac{d^3}{t^2}\right)
\sum_{s_1 v_1+s_2 v_2=n/d}\hspace{-.5cm}v_1 v_2 (v_1+v_2)\min(s_1,s_2).
\end{equation}

\subsection*{c. Calculation of $\kdz$}
The evaluation of (\ref{k0total}) will require to introduce the functions
\begin{equation}
f(n)=\frac{\mu(n)}{n}\ \ \ \ \ \ \ \textrm{and}\ \ \ \ g(n)=\sum_{d|n}\frac{\mu(d)}{d^2}
\end{equation}
For $f_1$ and $f_2$ two arithmetic functions, the Dirichlet convolution is defined by
\begin{equation}
f_1*f_2(n)=\sum_{d|n}f_1(d)f_2(\frac{n}{d}).
\end{equation}
Replacing the expressions found for $G_{q/d}$ and $H_{q/d}$ in Equation (\ref{k0total})
we get
\begin{eqnarray}
\label{deuxtermes}
\kdz&=&\frac{1}{q \nnpq}\sum_{d|q}(f*\mu)(d)d^3
\left\{\sum_{s_1 u_1+s_2 u_2=q/d}\hspace{-.3cm}\left(\frac{q}{2d}\right)u_1 u_2 (u_1+u_2)\min(s_1,s_2)
\right.\nonumber\\
&+&\left.\sum_{d'|q/d}d'^3g(d')
\hspace{-.5cm}\sum_{s_1 u_1+s_2 u_2=q/dd'}\hspace{-.5cm}u_1 u_2 (u_1+u_2)\min(s_1,s_2)\right\}.
\end{eqnarray}
Rewriting the constant $N_q$ given by (\ref{nqp}), using the inclusion-exclusion principle
and following the steps from Equations (\ref{stari}) to (\ref{starf}), we get
\begin{equation}
N_q=\sum_{d|q}(f*\mu)(d)d^2
\hspace{-.5cm}\sum_{s_1 u_1+s_2 u_2=q/d}\hspace{-.5cm}u_1 u_2 (u_1+u_2)\min(s_1,s_2).
\end{equation}
We see that the first 
term in (\ref{deuxtermes}) is equal to 1/2. If we set $\delta=d d'$, the second term gives
\begin{equation}
\label{2emeterm}
\frac{1}{q \nnpq}\sum_{\delta|q}(\frac{\delta}{d})^3\left(\sum_{d|\delta}(f*\mu)(d)g(\frac{\delta}{d})\right)
\sum_{s_1 u_1+s_2 u_2=q/\delta}\hspace{-.5cm}u_1 u_2 (u_1+u_2)\min(s_1,s_2).
\end{equation}
But
\begin{equation}
\sum_{d|\delta}(f*\mu)(d)g(\frac{\delta}{d})=[(f*\mu)*g](\delta)=[f*(\mu*g)](\delta)
\end{equation}
by associativity of Dirichlet convolution, and
\begin{equation}
(\mu*g)(\delta)=\frac{\mu(\delta)}{\delta^2}
\end{equation}
by Moebius inversion formula. Finally the term (\ref{2emeterm}) simplifies to $1/q$,
which completes the proof.

%%%%%%%%%%%%%%%%%%%%%%%%%%%%%%%%%%%%%%%%%%%%%%%%%%%%%%%%%%%%%%%%%%%%%%%%
%                  REFERENCES BIBLIOGRAPHIQUES
%%%%%%%%%%%%%%%%%%%%%%%%%%%%%%%%%%%%%%%%%%%%%%%%%%%%%%%%%%%%%%%%%%%%%%%%

\end{document}